# Solid-State Core-Exciton Dynamics in NaCl Observed by Tabletop Attosecond Four-Wave Mixing Spectroscopy

James D. Gaynor,[1,2] Ashley P. Fidler,[1,2,4] Yen-Cheng Lin,[1,2] Hung-Tzu Chang,[1] Michael Zürch,[1] Daniel M. Neumark,[1,2]* Stephen R. Leone[1,2,3]*

[1]*Department of Chemistry, University of California, Berkeley, CA. 94720, USA.*

[2]*Chemical Sciences Division, Lawrence Berkeley National Laboratory, Berkeley, CA. 94720, USA.*

[3]*Department of Physics, University of California, Berkeley, CA. 94720, USA.*

[4] Currently: *Department of Chemistry, Princeton University, Princeton, NJ. 08544, USA.*

*\*Authors to whom correspondences may be addressed:* (S.R.L) srl@berkeley.edu; (D.M.N.) dneumark@berkeley.edu

**Abstract**

Nonlinear wave-mixing in solids with ultrafast x-rays can provide new insight into complex electronic dynamics of materials. Here, tabletop-based attosecond noncollinear four-wave mixing (FWM) spectroscopy using one extreme ultraviolet (XUV) pulse from high harmonic generation and two separately timed few-cycle near-infrared (NIR) pulses characterizes the dynamics of the $Na^+$ $L_{2,3}$ edge core-excitons in NaCl around 33.5 eV. An inhomogeneous distribution of core-excitons underlying the well-known doublet absorption of the $Na^+$ $\Gamma$-point core-exciton spectrum is deconvoluted by the resonance-enhanced nonlinear wave-mixing spectroscopy. In addition, other dark excitonic states that are coupled to the XUV-allowed levels by the NIR pulses are characterized spectrally and temporally. Approximate sub-10 femtosecond coherence lifetimes of the core-exciton states are observed. The core-excitonic properties are discussed in the context of strong electron-hole exchange interactions, electron-electron correlation, and electron-phonon broadening. This investigation successfully indicates that tabletop attosecond FWM spectroscopies represent a viable technique for time-resolved solid-state measurements.

# I. INTRODUCTION

The explosive development of ultrafast nonlinear spectroscopy with extreme ultraviolet (XUV) and X-ray light pulses from both tabletop high harmonic generation (HHG) and X-ray free-electron lasers (XFELs) is rapidly advancing our knowledge of atoms, molecules, and materials [1]. XFELs now produce sub-100 femtosecond (fs) light pulses at sub-nanometer wavelengths with intensities up to $10^{20}$ W cm$^{-2}$ [2]. The high pulse intensities lend well to exploring many long-predicted nonlinear pulse sequences of multiple XUV / X-ray fields [3], such as X-ray stimulated Raman based methods [4], multidimensional X-ray experiments [5], soft X-ray second harmonic generation [6], and four-wave mixing (FWM) spectroscopy with X-rays [7]. By comparison to XFELs, HHG sources produce XUV and soft X-ray pulses of broader coherent spectral bandwidth that supports attosecond temporal resolution – characteristics that are ideal for capturing electronic dynamics that typically occur at <10 fs timescales [1,2]. The flexibility and autonomy of tabletop XUV sources in a modern optics laboratory also allows complicated nonlinear experiments and pulse sequences to be developed and quickly refined, such as those requiring multiple optical and near-infrared (NIR) pulses in noncollinear beam geometries [8,9].

Using XFELs, condensed matter physicists have exploited the high-flux <50 nm XUV wavelength pulses in transient-grating experiments with two time-coincident XUV pulses to form a spatially periodic excitation profile followed by a third pulse to measure electric susceptibilities and phonon propagation on length scales of several nanometers [10-13]. Using HHG sources, attosecond noncollinear FWM spectroscopy has measured electronic, vibrational, and vibronic dynamics in gas-phase atoms and molecules, characterizing excited state valence and Rydberg wavepackets and the influence of nearby dark states with sub-fs temporal resolution and elemental selectivity [8,9,14,15]. Key advantages of the attosecond FWM experiment include: signal enhancement through three resonant transitions between core-level states and neighboring states, background free detection of the XUV signal due to a noncollinear beam geometry, sub-fs synchronization with an array of optical pulses that can be combined with the XUV, and sub-fs time-domain tracking of core-excited state evolution through double-quantum coherence transition pathways. The success of attosecond FWM experiments in atomic and molecular physics suggests that this methodology can advance investigations of atom-like core-excitonic phenomena in solid-state materials, as well. We connect the atomic and the solid-state physics communities in this paper by implementing tabletop attosecond FWM spectroscopy to successfully measure core-level excited state phenomena of the solid-state insulator NaCl at the Na$^+$ L$_{2,3}$ edge near 33.5 eV.

Alkali halides are insulators with important electrical and optical properties arising from a large energy gap between their valence and conduction bands [16]. They are the simplest ionic solids, consisting of equal numbers of monovalent cation-anion species. Decades of investigations on the short-range interactions of bound electron-core-hole pairs in these materials (core-level excitons or "core-excitons") have aimed to characterize the nature of core-exciton states formed in alkali halides, such as NaCl [17-29] where some are associated with Na$^+$ and some with Cl$^-$. Weak dielectric screening inherent to wide gap insulators enhances electron-hole attraction; the highly localized electronic structure of the core-exciton approaches that of a Frenkel exciton [16,17,26,30]. In this highly localized limit, similarities to the electronic properties of atomic orbitals can aid the understanding of core-excitonic phenomena [31,32]. The elemental specificity of XUV and X-ray spectroscopy is ideal for probing these very localized electronic dynamics that reflect complex interactions between electrons and nuclei. Moreover, nonlinear spectroscopies using XUV pulses have the potential to unravel information, such as inhomogeneous broadening, that is typically hidden in linear absorption based methods [7,33] while exploiting this elemental specificity.

The class of third-order nonlinear techniques, which includes XFEL-based transient-grating experiments and attosecond FWM spectroscopy, specifies each of three incident fields by unique wavevectors to account for the emitted signals as the fourth wave. Attosecond FWM spectroscopy differs by temporally resolving the system evolution between all three pulses and by resonantly driving double-quantum coherence pathways used to generate the signal. The high temporal resolution achieved in the tabletop attosecond FWM spectroscopies is crucial for measuring dynamics on sub-10 fs timescales. The few-fs relaxation timescales measured for core-excitonic state dynamics are faster than the pulse durations typically produced thus far at XFELs, although shorter pulses may be expected soon. Here, attosecond FWM spectroscopy is applied to characterize the highly localized Na$^+$ core-exciton measured at the Na$^+$ L$_{2,3}$ absorption edge on the sub-10 fs timescale. We find that several more Na$^+$ core-excitons underlie the Na$^+$ L$_{2,3}$ absorption than previously reported, and that there are very different temporal dynamics that can be observed on such timescales. Distinct spectral and temporal behaviors of the time-resolved XUV emission are measured for both XUV bright states and a subset of XUV dark states that are uniquely probed with this method. The nonlinear nature and phase-matching dependence of the attosecond FWM experiment enables us to show that the well-known NaCl core-exciton absorption arises from a heterogeneous distribution of at least five core-exciton states.

## II. RESULTS

### A. The extreme ultraviolet spectrum of NaCl

The linear absorption spectrum of a 50 nm thick NaCl microcrystalline film from 32.5-37.5 eV is shown in Fig. 1. The strong absorption features in the 33.2-33.8 eV region arise from the absorption to core-excitonic states – bound electron-hole pairs involving a core-hole – at the Na$^+$ L$_{2,3}$ edge. From the atomic perspective, this is the excitation of a *2p* electron resulting in the Na$^+$ *2s$^2$2p$^5$3s* electronic configuration [17,18,30]. From a band structure perspective, the resemblance between the $Na^+$ $L_{2,3}$ edge spectrum and the UV absorption spectrum has resulted in the XUV features being ascribed to core-excitons at the Γ point in the Brillouin zone [17]. Three peaks comprise the lowest energy absorption band in Fig. 1 (inset), one weaker at 33.27 eV and two stronger, overlapping peaks at 33.42 eV and 33.61 eV. In previous literature the predominant doublet structure between one strong higher energy feature and one weaker lower energy feature is widely attributed to spin-orbit splitting, as is observed in the fundamental UV absorption spectrum [17]. To our knowledge, the partial splitting of the strong higher energy feature is not reported in the XUV absorption literature but such splittings have been observed for valence-excitons in the UV range and are considered further in terms of lattice imperfections and surface states in the discussion section. Photoelectron measurements of the $Na^+$ $2p$ core-exciton have shown XUV spectra with greater width than expected and a similar shoulder peak structure to the absorption spectrum in Fig. 1(inset) which are suggested, but not confirmed, to be due to surface states [34]. Due to the *p*-orbital symmetry of the core-hole, a triply degenerate state is expected in the absence of any energetic splitting. The specific origin of the weak band spanning 34-36 eV is not widely agreed upon, but it has been ascribed to transitions into bound electron-hole states lying just below the onset of the conduction band [18,19]. The peak centered at 36.5 eV corresponds to excitonic formation at a different point in the Brillouin zone [18,20], which is energetically similar to the $Na^+$ free ion $2s^22p^53p$ configuration [30]. The ~0.4 OD continuous absorption background is due to valence band transitions into high-lying conduction band states [17,18].

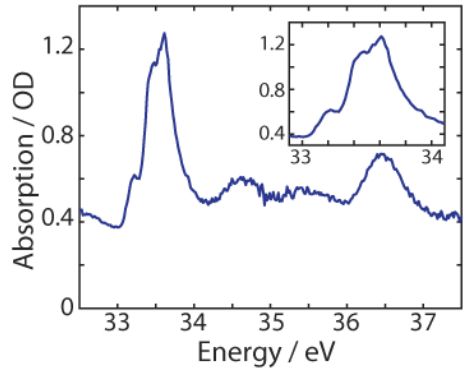

FIG. 1. Linear XUV Absorption Spectrum of NaCl around $Na^+$ $L_{2,3}$ Edge. The linear absorption spectrum of a NaCl thin film (50 nm) used in the FWM experiments of the $Na^+$ $L_{2,3}$ edge. The inset zooms in on the $Na^+$ core-exciton absorption features of interest in this study.

B. Attosecond XUV-NIR four-wave mixing spectroscopy of NaCl core-excitons

The schematic in Fig. 2(a) illustrates the principle of the attosecond FWM experiment. An attosecond pulse train ($k_{XUV}$) consisting of 2-3 sub-femtosecond bursts in the XUV spanning 25-45 eV is focused through the aperture of a silver-coated annular mirror onto the NaCl microcrystalline thin film sample. Two independently controlled NIR beams ($k_{NIR1}$, $k_{NIR2}$), spanning 550-950 nm (1.3-2.25 eV), are each routed through translational delay stages and then spatially overlapped with the XUV beam on the NaCl sample by reflection with the annular mirror. The delay stages controlling the XUV-NIR time delay ($\tau_1$) and the $NIR_1$-$NIR_2$ time delay ($\tau_2$) are arranged in series to optimize timing precision and stability in the time-dependent experiments discussed below.

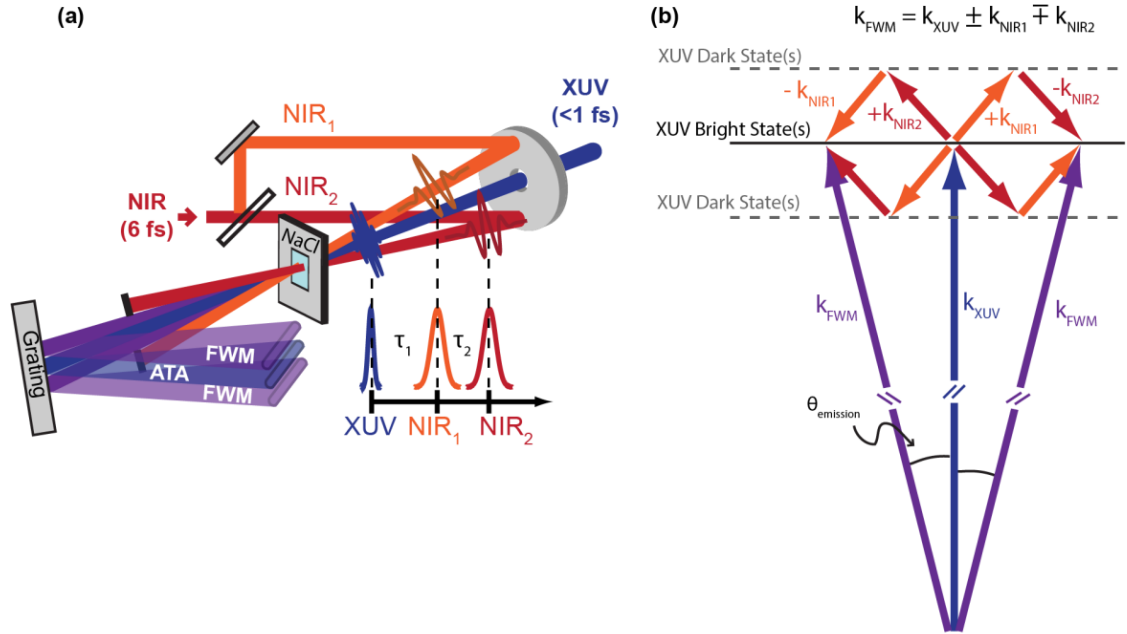

FIG. 2. Attosecond XUV-NIR Four-Wave Mixing Spectroscopy in a Noncollinear Beam Geometry. (a) Schematic of the beam geometry and pulse sequence. Two few-cycle NIR beams ($k_{NIR1}$/$NIR_1$, orange; $k_{NIR2}$/$NIR_2$, red) are vertically stacked and crossed with an attosecond XUV pulse train ($k_{XUV}$/XUV, blue) on the NaCl sample. The delays between each beam are independently tunable; the XUV-NIR delay is denoted $\tau_1$ and the delay between the two NIR pulses is $\tau_2$. The wave-mixing signals ($k_{FWM}$) are emitted at an angle, $\theta_{emission}$, and the attosecond transient absorption signals are collinear with the XUV pulse. (b) The initially excited XUV-bright states are coupled through XUV-dark states and back to the bright state manifold through the two sequential NIR interactions. The phase-matching condition determines the emitted signal direction. Controlling the pulse timing determines whether the time-resolved XUV emission is determined by evolution of the XUV-bright states or the XUV-dark states.

As shown in Fig. 2(b), the broadband XUV pulse excites the core-excitons at the $Na^+$ $L_{2,3}$ edge. One NIR interaction induces transitions from the core-excitonic excited states to a nearby XUV-dark state manifold; the second NIR interaction transfers the excited system back to an XUV-allowed manifold of core-excited states. The detected signals arise from the XUV emission of these excited states back to the ground state. Following the sample, an Al foil (0.15 µm thick) filters out the NIR beams and transmits the generated XUV signals, which are spectrally dispersed and imaged onto an X-ray charge-coupled device (CCD) camera. A more complete experimental description is published elsewhere [8,9] and details are in Appendix A.

Each of the $i^{th}$ fields in the attosecond FWM experiment can be expressed as $\mathbf{E}_i(t) = A_i(t-t_i)e^{-i(\omega_i(t-t_i)+\phi_i)}e^{-i\mathbf{k}_i \cdot \mathbf{r}}$ with a temporal profile $A_i(t-t_i)$ centered at $t_i$, and a carrier frequency $\omega_i$, phase $\phi_i$, and a unique wavevector $\mathbf{k}_i$ directionally projected onto a laboratory frame coordinate vector, $\mathbf{r}$. The noncollinear beam geometry exploits the phase-matching condition

$$\mathbf{k}_{FWM} = \mathbf{k}_{XUV} \pm \mathbf{k}_{NIR1} \mp \mathbf{k}_{NIR2} \quad (1)$$

during the sequential light-matter interactions *via* the third-order susceptibility, $\chi^{(3)}$, of the sample [33]. The momentum conservation requirement of Equation (1) results in the emission of spatially distinct XUV wave-mixing signals that can be detected background-free [8,9,14,15]. The FWM emission angle, $\theta_{emission}$, is easily calculated from the photon energies of the incident pulses ($\omega_{XUV}$, $\omega_{NIR1}$, $\omega_{NIR2}$) and the crossing angles between the NIR beams and the XUV beam according to the relation:

$$\theta_{emission} = \frac{\omega_{NIR1}\theta_1 \times \omega_{NIR2}\theta_2}{\omega_{XUV}} \quad (2)$$

where $\theta_1$ ($\theta_2$) is the angle between $\mathbf{k}_{XUV}$ and $\mathbf{k}_{NIR1}$. The chosen beam geometry results in vertically displaced wave mixing signals with respect to the XUV pulse such that $\mathbf{k}_{XUV}$ is centered on the X-ray CCD camera at $\theta_{emission} \approx 0$ mrad and the grating spectrally disperses in the horizontal dimension. As implied in Fig. 2A, the attosecond transient absorption (ATA) signal is also generated from two interactions with the same NIR pulse (e.g., $\mathbf{k}_{ATA} = \mathbf{k}_{XUV} + \mathbf{k}_{NIR1} - \mathbf{k}_{NIR1}$), emitted collinearly with $\mathbf{k}_{XUV}$, which can be simultaneously measured with the FWM signals.

At temporal overlap of all three incident pulses ($\tau_1 = 0$, $\tau_2 = 0$), many different transition pathways are possible. The FWM pathways relevant to this discussion are illustrated in Fig. 2B, all of which require initial excitation by $\mathbf{k}_{XUV}$. In principle, transition pathways including XUV-dark states that are within the NIR bandwidth are allowed; they may be either higher or lower in energy than the initially excited XUV-bright states.

The decay dynamics of the XUV-excited states can be measured by keeping NIR$_1$ and NIR$_2$ temporally overlapped with each other ($\tau_2 = 0$) and scanning the XUV-NIR delay ($\tau_1$); we refer to this as a "bright state scan". In a bright state scan, the phase-matching condition permits FWM signals with equivalent information content to be emitted at positive $\theta_{emission}$ (above $\mathbf{k}_{XUV}$) and negative $\theta_{emission}$ (below $\mathbf{k}_{XUV}$) angles. Additionally, the effect of the coherently excited dark states on the XUV emission can be measured by setting $\tau_1 = 0$ while $\tau_2$ is scanned; we refer to this as a "dark state scan". The dark state scan effectively drives a two-photon excitation (XUV ± NIR) into any XUV-dark state that has an appreciable transition dipole moment connecting the bright and dark state manifolds. The evolution of these dark states can then be monitored by scanning $\tau_2$. Notably, the information symmetry between the FWM signals measured at $\pm \theta_{emission}$ is broken in the dark state scan as soon as $\tau_2$ is beyond pulse overlap due to the phase-matching condition. Provided there are dark states accessible at lower and higher energies relative to the XUV-bright states, the selective measurement of the dynamics within two different dark state manifolds is possible by targeting the collection of FWM signal at either positive or negative $\theta_{emission}$. As shown below for the NaCl core-excitons at the $Na^+$ $L_{2,3}$ edge, the dark state scans can also help to narrow down the set of states participating in the bright state transition pathways.

The $Na^+$ core-exciton states observed in the XUV spectrum (Fig. 1) are measured using attosecond FWM spectroscopy and shown in Fig. 3(a) at temporal overlap of all three pulses ($\tau_1 = 0$, $\tau_2 = 0$). The FWM emission is centered at $\theta_{emission} \approx \pm 2.2$ mrad, in agreement with the expected divergence angles of $\theta_{emission} \approx \pm 1.4$-$2.4$ mrad for NIR crossing angles ($\theta_1$, $\theta_2$) of 18 mrad; the beam geometries are illustrated above the spectra. The differential absorption is plotted as $\Delta A = -log(I/I_0)$, where $I$ is the signal collected with all beams incident on the sample and $I_0$ is the signal collected when both NIR beams are blocked by an automated shutter. As mentioned above, the ATA signals can be collected simultaneously with the FWM signals due to the different angular dependences of the emissions. The ATA phase matching conditions result in emitted signals collinear with $\mathbf{k}_{XUV}$ at $\theta_{emission}$ centered around 0 mrad. The angular dispersion of the ATA signal in the $-1 < \theta_{emission} < 1$ mrad range is dependent upon the generation and focusing conditions of $\mathbf{k}_{XUV}$. The FWM signals are homodyne-detected, whereas the ATA signals utilize the collinear $\mathbf{k}_{XUV}$ as an intrinsic local oscillator for self-heterodyned detection [8]. As a result, bleach/emission (negative) and induced absorption (positive) features are distinguished in the ATA differential absorption.

Each of the two noncollinear NIR beams required for the FWM experiment generates ATA signal. Thus, the ATA signal measured in Fig. 3A is the sum of two equivalent ATA signals - one from each NIR beam. To confirm that the observed signals above and below the ATA region result from the FWM nonlinear process involving $\mathbf{k}_{XUV}$, $\mathbf{k}_{NIR1}$, and $\mathbf{k}_{NIR2}$, a spectrum was collected with $\mathbf{k}_{NIR1}$ blocked and is shown in Fig. 3B. As expected, the FWM signals observed at $|\theta_{emission}| > 1$ mrad vanish and the ATA signal measured at $-1 < \theta_{emission} < 1$ mrad is approximately half of the intensity of that measured in Fig. 3A. We note that the ATA signals often require lower CCD exposure times to avoid pixel saturation from $\mathbf{k}_{XUV}$ affecting the differential absorption measurement. Thus, the ATA signals were collected with 0.7 s CCD exposure and the FWM signals were collected with 2.0 s exposure resulting in the FWM signals effectively enhanced relative to the ATA signals by a corresponding factor of $2.0/0.7 \cong 2.86$.

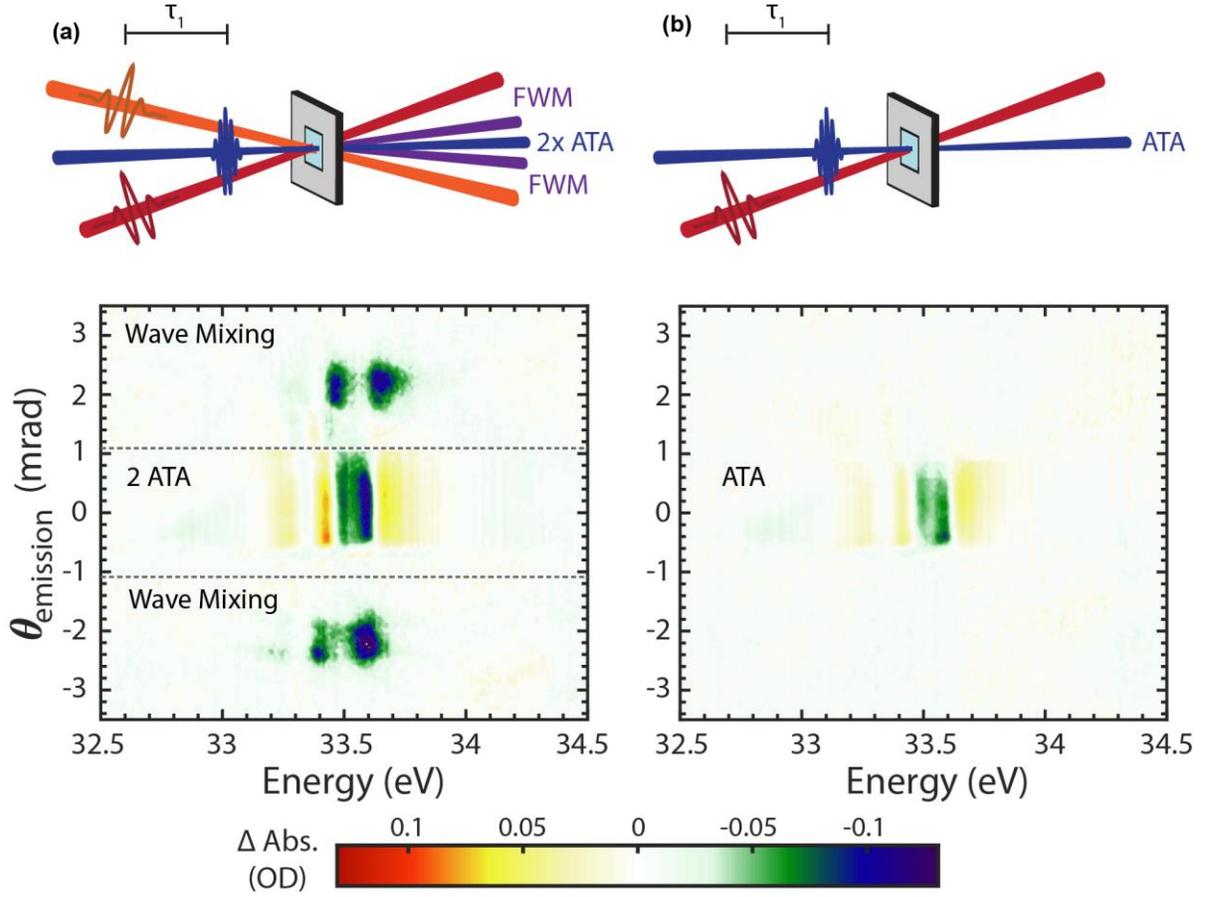

FIG. 3. XUV-NIR Nonlinear Wave Mixing Signal of $Na^+$ Core-Excitons in NaCl. (a) The $Na^+$ $L_{2,3}$ core-exciton signal observed under the FWM experimental conditions (all beams unblocked, pulse sequence above) at temporal overlap ($\tau_1 = \tau_2 = 0$). The background free FWM signals are observed at $|\theta_{emission}| > 1$ mrad while the ATA signals are found in the range $-1 < \theta_{emission} < 1$ mrad; negative $\Delta OD$ indicates emission. (b) Confirmation of the nonlinear wave-mixing nature of the detected signals from blocking $k_{NIR1}$; only the ATA signal from $k_{NIR2}$ is observed as expected. To avoid detector saturation, the ATA signals are collected with 0.7 seconds integration while the FWM signals are collected with 2.0 seconds integration, resulting in the FWM signals shown in (A) effectively enhanced relative to the ATA signals by ~2.86.

The magnitude of the strongest FWM signal from the core-exciton at 33.61 eV is plotted as a function of NIR fluence in Fig. 4(a). The linear dependence on energy fluence with each NIR pulse gives additional confirmation of the FWM nature of the reported signals [12]. Higher order signals (e.g., six-wave mixing) are expected to be negligible as signal magnitudes decrease for each successively higher electric field dependence [33]. The three most intense core-exciton features at temporal overlap are measured to have different FWM signal generation efficiencies, defined as the ratio of the integrated intensities of the FWM signal magnitude ($I_{FWM}$) and the incident XUV pulse ($I_{XUV}$), $I_{FWM}/I_{XUV}$. Figure 4(b) presents the FWM efficiency for the excitons at 33.27 eV (black), 33.42 eV (light blue), and 33.61 eV (green) as a function of $NIR_1$ fluence. A linear increase for all three core-excitons up to 12 mJ/cm² of NIR fluence is evident and the data for each core-exciton is best fit to a straight line up through this value (solid lines). However, the core-excitons at 33.26 eV and 33.42 eV behave nonlinearly with plateaus in the FWM efficiencies for $NIR_1$ fluences above 12 mJ/cm$^2$, as shown by the quadratic polynomial fits (dashed lines). There is a negligible difference between the linear and quadratic fits for the 33.61 eV core-exciton through NIR fluences of 25 mJ/cm$^2$. A possible explanation for the varied FWM efficiencies with increasing NIR fluence is that the three core-exciton states utilize different dark states in the wave-mixing process. For example, the two core-excitons with saturating efficiencies at high NIR fluences could involve dark states that are more susceptible to multiphoton absorption effects. Aside from the fluence dependence measurement, all experiments reported here were performed with NIR pulse fluences of 8-10 mJ/cm$^2$ to minimize undesirable multiphoton effects.

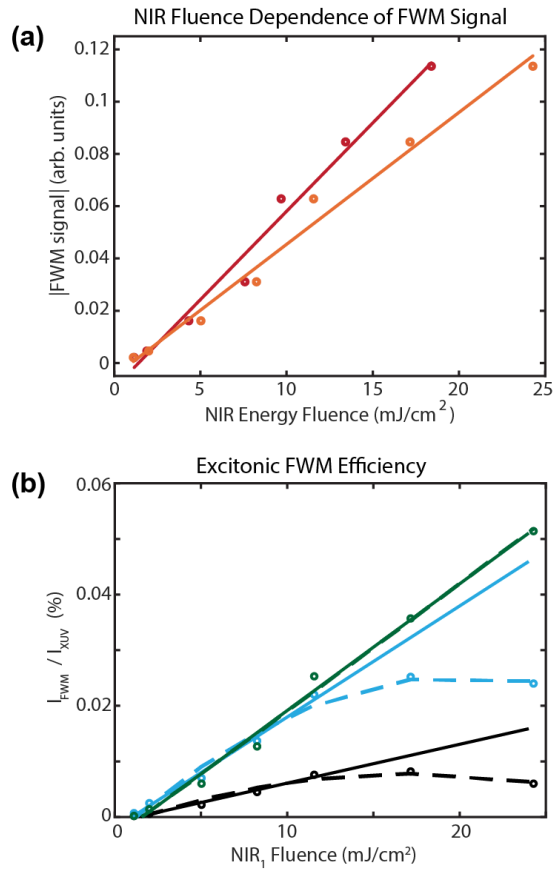

FIG. 4. NIR Fluence Dependence and Generation Efficiency of the FWM Signal. (a) The NIR fluence dependence of the strongest Na$^+$ core-exciton at 33.61 eV for $\mathbf{k}_{NIR1}$ (orange) and $\mathbf{k}_{NIR2}$ (red); spectra are collected at temporal overlap. (b) The excitonic FWM efficiency calculated for the three main core-excitons with increasing $\mathbf{k}_{NIR1}$ fluence observed at temporal overlap. Linear (solid) and quadratic (dashed) fits; 33.27 eV = black, 33.42 eV = light blue, 33.61 eV = green. Fluence dependence of FWM signal and excitonic efficiencies are all linear below 12 mJ/cm$^2$, NIR fluences in the 8-10 mJ/cm$^2$ were used in the FWM studies to stay in the linear regime.

C. Core-exciton dynamics resonantly probed with FWM spectroscopy

In the bright state scan configuration, the transient FWM signals track the temporal evolution of the core-excitonic wavepacket that is excited by $\mathbf{k}_{XUV}$. The time-dependence of the Na$^+$ core-exciton states is measured in the bright state scan shown in Fig. 5(a). The transient spectrum is obtained by plotting the integrated FWM signal ($-2 \geq \theta_{emission} \geq -2.4$ mrad) as $\tau_1$ is scanned. While the same information is present in the FWM signal at positive $\theta_{emission}$ in the bright state scan, the signal at negative emission angles was easier to isolate experimentally from the ATA signal and the residual XUV scattered light by using a combination of irising and filtering upstream in the HHG beamline (see Appendix A). Coherence decay times of $7.1 \pm 0.9$ fs and $6.7 \pm 0.8$ fs were determined for the bright state scan line outs at 33.42 eV and 33.61 eV, respectively, by fitting the traces to a single exponential decay convolved with Gaussian instrument response functions of $5.2 \pm 0.9$ fs and $5.1 \pm 0.8$ fs. The lowest energy feature centered at 33.27 eV during temporal overlap is weaker in magnitude and appears to have comparable lifetimes to the higher energy peaks but undergoes an ~0.08 eV blueshift over the initial 20 fs of core-exciton relaxation.

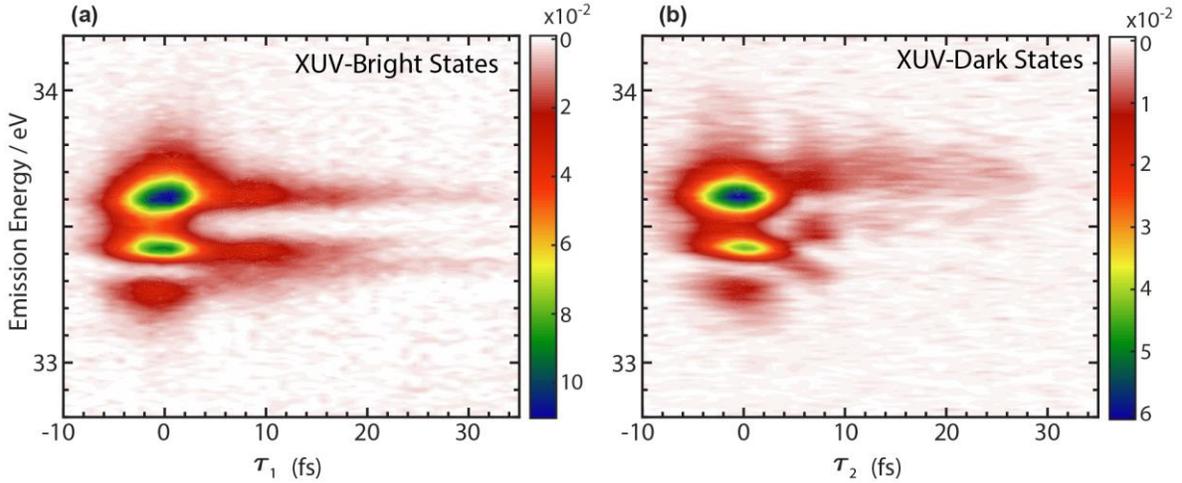

FIG. 5. Time-Dependent FWM Signals Measure Excitonic Dynamics. (a) Time-resolved XUV emission of the FWM signal ($-2.0 > \theta_{emission} > -2.4$ mrad) collected as a function of XUV-bright state evolution ($\tau_1$ scanned, $\tau_2 = 0$). (b) Time-resolved XUV emission of FWM signal collected as a function of XUV-dark state evolution following initial two photon ($\mathbf{k}_{XUV} + \mathbf{k}_{NIR1}$) excitation into the XUV-dark states ($\tau_1 = 0$, $\tau_2$ scanned).

In the dark state scan configuration, the initial two-photon interaction ($\mathbf{k}_{XUV} \pm \mathbf{k}_{NIR1}$) excites a wavepacket propagating in XUV-forbidden states that are within the few-cycle NIR bandwidth ($\omega_{XUV} \pm \omega_{NIR1}$). The dark state wavepacket is monitored by plotting the same integrated range of $\theta_{emission}$ as in the bright state scan for each $\tau_2$ delay of $\mathbf{k}_{NIR2}$. Figure 5(b) shows the transient FWM dark state signal corresponding to the phase-matched transition pathways with $\mathbf{k}_{FWM} = \mathbf{k}_{XUV} + \mathbf{k}_{NIR1} - \mathbf{k}_{NIR2}$ momentum conservation after pulse overlap; this is a consequence of the geometric orientation of the pulses and the temporal causality enforced by the relative pulse timings. That is, since $\mathbf{k}_{NIR1}$ is aligned above $\mathbf{k}_{XUV}$ prior to the sample, the phase-matched

FWM signals from the $\mathbf{k}_{XUV} + \mathbf{k}_{NIR1}$ interaction will appear at negative $\theta_{emission}$, which is the signal of interest here. The dark states involved most strongly with all the measured FWM signals can thus be determined to be greater in energy than the XUV-excited Na$^+$ core-excitons due to the minimal time-dependent FWM signal of dark state scans at positive $\theta_{emission}$ (see Supplementary Material, Fig. SM 1.) [35].

The spectral profile of the dark state scan significantly differs from the bright state scan in Fig. 5. While the core-excitons measured at $\tau_2 = 0$ have the same peak positions as the bright state scan (33.27, 33.42, and 33.61 eV), the peak positions shift to 33.34 eV, 33.47 eV, and 33.67 eV by $\tau_2 = 7$ fs in the transient dark state scan. The decay of the 33.67 eV peak can be fit to a single exponential yielding an 8.1±1.0 fs coherence decay. It is difficult to retrieve meaningful decay times from the lower energy peaks without invoking a more complex kinetic model because of the different spectral profile observed at 7 fs in comparison to that at temporal overlap.

A spectral fitting of the FWM signal at temporal overlap is shown in Fig. 6. A snapshot of the core-excitons accessible upon initial excitation is determined by fitting the FWM spectrum at temporal overlap to the states observed in the bright state scan (Fig. 6(a)) and then comparing the residuals to the core-excitons measured independently in the dark state scan at 7 fs (Fig. 6(b)). The inhomogeneous broadening of the excitons is evident when Gaussian distributions are used to provide the best fit to the line shapes in Fig. 6(a). The peak fits in Fig. 6(a) are centered at the bright state excitonic energies 33.27 eV, 33.42 eV, and 33.61 eV, which have FWHM linewidths of 0.050 eV, 0.066 eV, and 0.098 eV, respectively. The underlying background absorption has been subtracted and the spectra are normalized to the 33.61 eV peak for fitting analysis. The peaks are fit with amplitudes of 0.29, 0.83, and 1.00, respectively. The core-excitonic spectrum from the dark state scan in Fig. 6(b) at $\tau_2 = 7$ fs is in very good agreement with the fit residuals for the two higher energy features centered at 33.47 eV and 33.67 eV, but less so for the lowest energy feature. The spectral resolution of these individual components demonstrated in Fig. 6 is a notable result demonstrating the power of these nonlinear experiments.

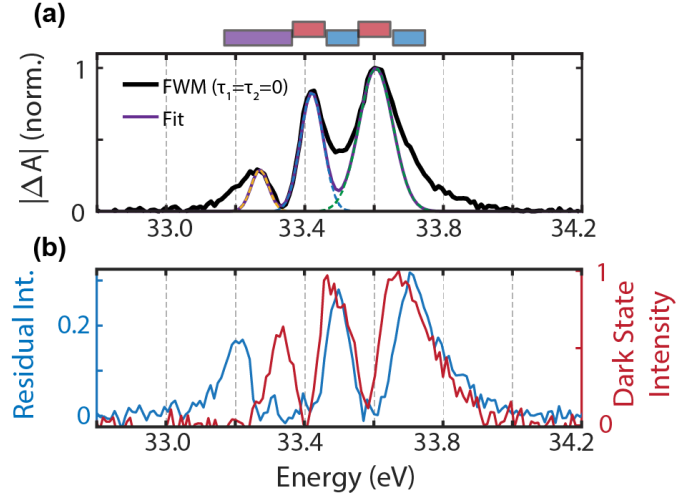

FIG. 6. Deconvolution of Core-exciton FWM signals. Spectral deconvolution of the time-zero spectrum with core-excitonic components identified in both the bright and dark state scans. (a) Temporal overlap FWM spectrum (black) fit with the three primary core-excitons identified in the bright state scans with Gaussian distributions and (b) the fit residuals (blue, left axis) compared to the core-excitons (normalized, red, right axis) identified in the dark state scans following overlap at a 7 fs delay. The blocks above the top plot highlight the five distinct core-excitonic regions discussed in the text. The red blocks refer to distinct states observed only in the bright state scan in (a); blue blocks refer to distinct states identified through the comparison of the residuals and dark state scans in (b) and the purple block on the low energy region highlights features observed in both (a) and (b) which may correspond to more than one distinct state.

## III. DISCUSSION

Previous interpretations of the core-excitonic spectra at the $Na^+$ $L_{2,3}$ edge in NaCl have relied heavily on the understanding of valence-excitonic spectra measured in the UV due to their close resemblance [17,18,21]. It is well known that the valence-exciton in NaCl consists of the excitation of a $Cl^-$ *3p* electron to a neighboring $Na^+$ *3s* orbital, which remains bound to the *3p* hole on the $Cl^-$ ion [21,22,29]. The electron is localized on the Na atom with *s*-orbital-like symmetry for both the valence-exciton and the core-exciton [17,22,23]. The core-exciton at the $Na^+$ $L_{2,3}$ edge is expected to have a smaller excitonic radius compared to the valence-exciton due to the hole also being confined to the Na *2p* orbital. The predominant doublet structure, reported widely in the linear absorption of both the valence-exciton and the core-exciton, is typically ascribed to spin-orbit splitting of the *p* orbitals [17,18,21,24]. However, we observe more distinct core-excitonic components through the spectral deconvolution of the $Na^+$ $L_{2,3}$ absorption edge using attosecond nonlinear FWM spectroscopy than is observed in the linear spectrum of Fig. 1(inset).

### A. Deconvolving inhomogeneous distributions of $Na^+$ core-excitonic states

The nonlinear wave-mixing spectroscopy used here exploits the dependence of the FWM signal on three incident electric fields to isolate different core-excitonic sub-ensembles within the inhomogeneous distribution of excitonic states present at the Na$^+$ L$_{2,3}$ edge. Importantly, this is achieved through resonantly driving the transitions between XUV-bright and XUV-dark states using the two noncollinear NIR pulses. Clear evidence that at least five distinct core-excitonic states underlie this absorption band is revealed by the transient FWM data in Fig. 5 and the spectral deconvolution in Fig. 6. It is well established from previous wave-mixing studies of atomic and molecular gas phase systems that different wave-mixing signals may contribute through Λ-type, V-type, and ladder-type transition pathways [8,9,15]. The combination of bright and dark state spectra with signal selectivity at positive and negative θ$_{emission}$ in Fig. 5 and Fig. SM 1 [35] confirms that pathways analogous to the Λ-type atomic transitions are the primary contributors to the observed signals. The experiments demonstrate that these dark states must be energetically situated within the bandwidth of one NIR photon greater in energy (within 1.3-2.2 eV) of the XUV excited core-excitons. They must also have *p*-orbital-like symmetry to carry non-negligible oscillator strength for the transition from the excited core-excitonic states with *s*-orbital symmetry localized on the Na$^+$ core.

The time-dependence of the bright state FWM signals in Fig. 5(a) distinguishes the nature of the core-excitons at 33.42 eV and 33.61 eV from the core-exciton at 33.27 eV. The two higher energy core-excitons decay monoexponentially on the few fs timescale without any change in spectral position or shape. There is possibly some evidence of a weak recurrence around τ$_1$ = 8 fs resembling a coherent oscillation in these two features, as well. These decays reflect the dephasing timescales of core-excitonic states with relatively stable energies since no significant shift in line shape is measured during the decay. A different dynamic is observed in the lowest energy core-exciton, which blue-shifts by ~0.08 eV over approximately 20 fs. While the origin of this dynamic is not yet certain, possible explanations include core-hole stabilization for a non-equilibrated core-exciton or excitonic wavefunction motion away from its initial excited state configuration on this timescale.

Recent attosecond studies of core-exciton dynamics in Si [31] and MgO [32] have alluded to the importance of dark excitons. Attosecond FWM spectroscopy has the unique ability to directly measure the effect of the dark states on the evolution of the initially excited ensemble of core-excitons. The spectral shifting observed in the NaCl dark state scans by 7 fs after temporal overlap (Fig. 5(b)) arises from the dark state excitonic wavepacket propagating away from the electronic distribution of initially excited states – analogous to an excited state molecular wavepacket moving away from the Franck-Condon region [9]. Following previous assignments of the Na$^+$ L$_{2,3}$ absorption to core-excitons at the Γ-point in the Brillouin zone [17], the observed dynamics would reflect the excitonic system propagating away from the Γ point. However, without knowledge of the specific dark states involved in the measured pathways, it is unclear to what extent the band picture can be applied. We note the discrepancy between the single particle band-picture and the highly localized excitonic picture here and emphasize that we cannot confirm the band-picture from our measurements.

To better understand the dark state scan data, consider that the nonlinear wave-mixing signal strengths are proportional to the product of the four transition dipole moments driven through the field-matter interaction, $\mu_{XUV}\mu_{NIR1}\mu_{NIR2}\mu_{FWM}$, where the subscripted labels correspond to those of the electric field wavevectors in the experimental pulse sequence (Equation 1). The transition dipole moments $\mu_{XUV}$ and $\mu_{FWM}$ represent allowed transitions between ground states and core-excitonic states (blue and purple arrows in Fig. 2(b)) while the $\mu_{NIR}$ represent transitions between the XUV-allowed core-excitonic states and the nearby dark states (red and orange arrows in Fig. 2(b)). Importantly, the $\mu_{NIR}$-dependence of the nonlinear signals probed here distinguishes the selection rules from those of the linear absorption spectrum, which lends the nonlinear techniques greater sensitivity to states within an inhomogeneous linear absorption spectrum. For this reason, there is no guarantee that the entire set of states measured in the linear absorption spectrum will be reproduced in the nonlinear signal emission, especially when the transition dipole moments connect states of such dramatically different relative energies (i.e., $\mu_{XUV}$ compared to $\mu_{NIR}$). The blue-shifted spectra in the dark state scans after temporal overlap (with the strongest signal at τ₂ = 7 fs) show that the evolution of some of the dark states coherently excited by the two photon $\mu_{XUV} + \mu_{NIR1}$ interaction results in the $\mu_{NIR2}$ dipole moments connecting a different set of core-excitons than in the bright state scan. The similarities and differences between the dark state spectrum and the fitted residuals in Fig. 6 support this interpretation.

It is important to note that the very measurement of the bright and dark state time-resolved spectra demonstrates that the bright core-excitonic states and the dark states maintain a coherent relationship for at least 10-20 fs. The double-quantum coherence pathways probed in this technique allow the direct measurement of coherence (dephasing) timescales (known as T₂) which are related to population lifetime (T₁) and environmentally-induced pure-dephasing (T₂*) of excited states in the well-known relationship: T₂ = 2T₁ + T₂*. In contrast to gas phase systems where negligible environmental pure-dephasing has been assumed, various sources of pure-dephasing can influence the spectral features and temporal dynamics in condensed phase systems as discussed in the following paragraphs.

The agreement between the FWM time zero spectrum and the core-excitons measured over 33.4-33.8 eV in the bright and dark state scans indicates that four distinct core-excitons are all accessible upon initial excitation in this energy range. By contrast, the lower energy range dynamics (33.0-33.4 eV) show that the core-exciton and the coupled dark states require temporal evolution of the excited wavepacket to acquire sufficient oscillator strength. The complex spectral dynamics in this range prove the existence of at least one more distinct core-exciton, though more than one state could be implicated from the lack of agreement to the spectral fitting in this range. This lack of agreement on the low energy end of the FWM spectral fit in Fig. 6 is not surprising. These excitonic states, presumed to be at the $\Gamma$ minimum in earlier studies, clearly shift, and perhaps undergo energetic splitting as in the dark state scan, on few-fs timescales. The complexity of the observed dynamics, especially at lower emission energies, is underscored by the blue-shifting feature in the bright state scans apparently converging to the same emission energy as observed from the dark state scan feature at $\tau_2 = 7$ fs. Collectively, at least five core-excitons have been identified at the $Na^+$ $L_{2,3}$ edge of NaCl through the attosecond FWM spectra reported here.

B. Physical insight into measured coherence decays: short-range electronic interactions

As shown above, higher-order nonlinear spectroscopies have an advantage over linear spectroscopies for deconvolving inhomogeneously broadened distributions [33]. It is important to contextualize these results with respect to previous literature discussing physical mechanisms for spectral splitting and broadening. Interestingly, the conclusions drawn from the FWM results above are further corroborated through the comparison between the linear absorption spectrum measured here and further details about the linear spectra elucidated in the literature. The $Na^+$ $L_{2,3}$ absorption spectrum of the core-exciton in Fig. 1 (inset) shows a doublet splitting of 0.255 eV using the centroid of the broader, higher energy feature (~33.52 eV), which is in exact agreement with the work of Nakai et al. [17]. The peak intensity ratio of 0.26 from Fig. 1 compares very well to the 0.21 ratio reported by Nakai et al. [17] for liquid nitrogen temperature spectra; the slightly lower ratio might be expected due to the narrowed line width and greater peak intensity of their 33.55 eV excitonic peak at low temperatures. As mentioned in the Results section, the partial splitting we observe for the two core-exciton states at 33.42 eV and 33.61 eV underlying the most intense absorption is not discussed in previous XUV literature, but they have been shown in the UV literature possibly due to electron-hole exchange correlation effects observed through lattice strain as measured by piezoreflectivity measurements [28] and changes in crystalline structure in similar alkali chlorides measured by UV absorption [21]. We observed the resolution of these features to diminish after constant irradiation on the same sample volume for extended periods due to the 33.42 eV peak reducing in amplitude. This suggests that the 33.42 eV core-exciton state could be very sensitive to lattice imperfections [21] and crystalline orientation [28] as characterized in the UV for the valence-excitons, or surface states as suggested in XUV photoelectron measurements [34]. Recent calculations of the $Na^+$ $L_{2,3}$ absorption of many sodium halides suggests that the core-exciton is sensitive to the crystal structure despite its high localization on the Na atom, requiring more detailed theoretical investigations of electronic structure and excitonic interaction on a peak-by-peak basis [36]. Thicker samples also diminished the resolution of the two larger peaks by substantially attenuating the main absorption feature. The difficulty of making and characterizing such thin films does not allow us to confirm single crystallinity of the studied samples (see Appendix B). As a result, we cannot claim that the FWM results, and the analysis above, apply exclusively to a single crystalline NaCl structure. However, these FWM results definitely reveal more features than the three discernible peaks in the linear absorption spectrum shown in Fig. 1 (inset) through the spectral deconvolution. While the doublet spectral shape of the linear absorption, consisting of the splitting between the peak at 33.27 eV and the two overlapped features making up the main peak with its centroid at 33.52 eV, is generally described through the lens of spin-orbit coupling, the following discussion points out that a refined description of the core-exciton line shapes and amplitudes is still debated and the alkali halide core-excitonic properties are not completely understood.

Spectra at liquid nitrogen temperature reported by Nakai *et al.* [17] show that the core-exciton peaks are much broader than the valence-exciton peaks and that the core-exciton line shapes cannot be described simply by electron-phonon broadening. While the large effective mass of the Na *2p* core-hole should enhance exciton-phonon coupling, the strong coupling regime is expected only when sufficient energy in the absorption bandwidth is available to excite a large number of phonons (e.g., > 10 phonons) [37]. The NaCl acoustic phonon spectrum spans 9-23 meV and the longitudinal-optical phonons are in the range of 24-40 meV [38-40]. The Gaussian line shapes observed here for $Na^+$ core-excitonic states (e.g., the spectral fits in Fig. 6) suggest that exciton-phonon coupling is not within the strong regime for the $Na^+$ core-excitonic states, as the peaks measured by FWM are too narrow to allow for a significant population of phonons – especially the longitudinal-optical phonons that typically drive exciton-phonon couplings. Interestingly, the results here suggest that exciton-phonon coupling in NaCl does not exert as much influence over the core-exciton as has been recently discussed for another ionic insulator, MgO, where exciton-phonon coupling decreases the core-exciton decay times considerably [32].

Auger decay processes for the $Na^+$ core-exciton in NaCl are not currently known well enough to fully account for the measured line broadening, either. Photoemission studies report average Auger lifetimes for the Na *2p* core-hole in NaCl of 30 meV (138 fs) with rather large uncertainty, permitting lifetimes up to 60 meV (69 fs) [41]. Nakai *et al.* [17] cite a 10 meV (414 fs) Auger decay and invoke electron-electron interactions between core-excitons and a continuous band of nearby states to describe the low-temperature core-exciton spectra. While the 66 meV and 98 meV bandwidths extracted for the core-excitonic states resolved at 33.42 eV and 33.61 eV are equivalent to 62 and 42 fs lifetimes, respectively, the time-domain FWM measurements here show faster coherence decays than expected from the currently reported Auger lifetime broadening. This is suggestive of more complex dephasing mechanisms at play for these core-excitonic states which will require in-depth theoretical investigations to determine fully.

The relative intensities of the doublet peaks are widely known to deviate from the 2:1 ratio expected by considering state multiplicity [18,20-23,25]. Onodera and Toyozawa [23] introduce exchange interaction between the electron and hole, mixing the excitonic states, to account for the different doublet peak intensity ratios. Using their model to analyze the linear absorption spectra, the electron-hole exchange interaction and spin-orbit interaction are extracted, respectively, as 0.0438 eV and 0.1149 eV for the valence-excitons, and 0.24 eV and 0.21 eV for the $Na^+$ core-excitons [17,23,24]. Applying the Onodera-Toyozawa model to the centroids of the general doublet structure in the 33.0-33.8 eV range of Fig. 1, our linear absorption agrees well with the previously reported exchange interaction and spin-orbit splitting for the $Na^+$ core-excitons [17,18,28]. This fact, alongside the proof from the FWM data (Fig. 5 and 6) that several core-excitons underlie this general doublet structure of the linear absorption spectrum, strengthens support for the existence of at least five distinct core-excitons at the $Na^+$ $L_{2,3}$ edge. Finally, greater peak multiplicities than the doublet for the alkali halide excitons have also been measured by reflectivity [25,28], transmission [21], low energy electron loss [27], and photoelectron spectroscopy [34] and predicted [26,29,36] to arise from the crystal field symmetry surrounding the alkali atom and varied crystalline orientations. Collectively, the above alkali halide exciton literature suggests that the prominent doublet structure may, in fact, be composed of an inhomogeneous distribution of core-excitons with a complex array of physical processes governing their properties. The FWM data reported here offer further evidence toward this end.

## IV. CONCLUSIONS

These experimental insights divulge a wealth of new information about the distributions of core-excitons in the prototypical ionic insulator, NaCl, studied here in polycrystalline 50 nm thick films. The localized nature of the electron and hole to the $Na^+$ ion suggests that the identified core-excitons with inhomogeneous line shapes result from additional short-range perturbations. A combination of strong electron-hole exchange interactions and electron-electron correlations likely yield spectral features with weak perturbations due to the likley poly-crystalline structure, augmented by broadening due to the phonon spectrum and site imperfections. An in-depth theoretical treatment is needed to visualize and more accurately describe the nature and dynamics of the core-excitons observed in the bright and dark state scans at the $Na^+$ $L_{2,3}$ edge. A complete picture of the physical mechanisms responsible for these features requires further theoretical developments to accurately and simultaneously treat spin-orbit coupling, electron-hole exchange correlation, and multielectron correlations.

Strong connections between well-studied phenomena in atomic physics and the exciton literature from the solid-state physics community have been established with this study. The implementation of nonlinear wave-mixing spectroscopy with XUV and NIR pulses described here is uniquely able to measure the effect of dark states on core-excitonic properties; time-dependent splitting of core-excitonic states are observed resulting uniquely from the evolution of XUV-dark states. We have shown direct evidence that the $Na^+$ core-exciton doublet spectrum is composed of several inhomogeneously broadened excitons with sub-10 fs coherence lifetimes. Our technique complements time-resolved transient-grating spectroscopies using XUV pulses generated at XFELs by exploiting resonant, phase-matched transitions between atomic-like states in a solid-state system to directly measure few-femtosecond dynamics in the time domain. By comparison to XFEL-based transient-grating experiments with high flux XUV pulses, the tabletop experiments discussed here directly access ultrafast, highly localized, atom-like electronic properties that are expected to occur at single-nm length scales and smaller. A new bridge between the attosecond atomic and molecular optics researchers and the solid-state physics community has been built, expanding the possibilities of future explorations in condensed matter and materials science that require atomic specificity and sub-fs temporal resolution.

## ACKNOWLEDGEMENTS

The authors thank Romain Géneaux for fruitful discussions of the data and interpretation. This work was performed by personnel and equipment supported by the Office of Science, Office of Basic Energy Sciences through the Atomic, Molecular and Optical Sciences Program of the Division of Chemical Sciences, Geosciences, and Biosciences of the U.S. Department of Energy at LBNL under Contract No. DE-AC02-05CH11231. Motivation for this work was provided by recently supported research through Air Force Office of Scientific Research (AFOSR), grant No. FA9550-20-1-0334. J.D.G. is grateful to the Arnold and Mabel Beckman Foundation for support as an Arnold O. Beckman Postdoctoral Fellow. A.P.F. acknowledges support from the National Science Foundation Graduate Research Fellowship Program. Y-C.L. acknowledges financial support from the Taiwan Ministry of Education. H.-T.C., M.Z., and S.R.L. acknowledge support from the AFOSR (FA9550-15-1-0037 and FA9550-19-1-0314) and the W.M. Keck Foundation (No. 046300).

J.D.G., A.P.F., Y-C.L., D.M.N., and S.R.L. designed the experiments; H.-T.C. and M.Z. observed the first attosecond transient absorption core-excitonic signals in NaCl, which formed the basis for the FWM experimental campaign; S.R.L. proposed the FWM experiments based on the previous observations of H.-T.C. and M.Z. using ATA spectroscopy. J.D.G., A.P.F., and Y-C.L. performed the FWM experiments; J.D.G. conducted the analysis; J.D.G., A.P.F., Y-C.L., H.-T.C., M.Z., D.M.N., and S.R.L. discussed results; J.D.G. wrote the paper.

APPENDIX A: METHODS – ATTOSECOND FOUR-WAVE MIXING APPARATUS

The output of a Ti:sapphire laser (Femtopower, 1 kHz repetition rate, 1.7 mJ/pulse, 22 fs, 780 nm) is spectrally broadened in a 2 meter long stretched hollow core fiber of 500 μm inner diameter (Few-Cycle Inc.) filled with 2.1 bar of neon gas. Seven pairs of double angled chirped mirrors (Ultrafast Innovations, PC70), fused silica wedge pairs, and 2 mm of ammonium dihydrogen phosphate (ADP) crystal are used to temporally compress the pulses to sub-6 fs durations, providing compressed broadband NIR pulses spanning 550-950 nm with ~600 μJ/pulse energies. The beam is then split by a 75:25 (R:T) beam splitter to separate the pulse used to create the XUV pulse ($\mathbf{k}_{XUV}$) from the pulse used to generate the two noncollinear NIR beams ($\mathbf{k}_{NIR1}$ and $\mathbf{k}_{NIR2}$). The reflected beam is focused by a concave silver mirror (f=50 cm) into a vacuum apparatus at $10^{-6}$ Torr containing a gas flow cell with krypton at ~4 Torr for HHG. The interaction of the few-cycle NIR field with the krypton gas generates an attosecond pulse train of 2-3 sub-fs XUV bursts in the 25-45 eV range. A vertically adjustable Al foil (0.15 μm thickness, Lebow) attenuates the co-propagating NIR field used to drive the HHG process to spectrally isolate the XUV pulse. The transmitted XUV pulse is refocused by a gold-coated toroidal mirror through the annular mirror into a NaCl thin film sample. The XUV intensity at the sample is estimated to be ~$10^{10}$ W cm$^{-2}$ on the sample due to the low HHG efficiency.

The transmitted pulse of the 75:25 beam splitter is delayed relative to the XUV beam using a piezoelectric translation stage (P-622 with E509 controller, Physik Instrumente) for control over $\tau_1$, and then the beam is split with a 50:50 beam splitter to form NIR$_1$ (reflection) and NIR$_2$ (transmission). Control over $\tau_2$ is achieved by routing NIR$_2$ onto a second piezoelectric translation stage. Two concave silver mirrors (f=1 m) then direct NIR$_1$ and NIR$_2$ above and below the annular mirror aperture, respectively, and spatially overlap them with the XUV beam on the sample. A finely adjustable iris is placed prior to the 50:50 beam splitter to systematically attenuate the noncollinear NIR beams. The NIR$_1$ (NIR$_2$) beam diameters at focus are characterized using a beam profiler (DataRay WinCamD) to be 300 μm (350 μm) at FWHM intensity for this work. A slight ellipticity of NIR$_2$ explains the relative trend in fluences shown in Fig. 4. Temporal and spatial overlap is determined by ATA signal of Ar gas (14.5 Torr) provided by a 1 mm gas cell accessible with a motorized sample translation stage. The NIR pulse envelope is estimated at 5.6 fs from the rise time of the Ar 3s4p autoionization signal measured by ATA (see Supplementary Material, Fig. SM 3) [35]. A vertically adjustable Al foil (0.15 μm, Lebow) attenuates the noncollinear NIR beams after the sample. The transmitted XUV light is spectrally dispersed in the horizontal plane by a gold-coated flat-field grating (01-0639, Hitachi) and recorded using a 1340 x 400-pixel X-ray CCD camera (Pixis XO 400-B, Princeton Instruments). The wave-mixing signals can be isolated and optimized using the two vertically adjustable filters, an automated iris placed right after the toroidal mirror, and a vertically translatable camera mount that allows for the unwanted ATA signals to be translated off of the CCD imaging area. Typically, the HHG alignment is optimized for photon flux and narrowest vertical intensity distribution for the XUV beam imaged on the CCD. Very fine adjustments of the filters and the iris are then used to spatially filter the vertical intensity profile of **k**$_{XUV}$ such that FWM signals at small θ$_{emission}$ can be well separated from overlap with ATA signals and residual scatter from **k**$_{XUV}$; this is increasingly necessary at higher XUV energies due to phase-matching angles. The bright and dark state scans were acquired with 10000 laser pulses for each time step and scanned with 1 fs steps. Each spectrum in the fluence dependent studies was acquired with 14000 laser pulses at temporal overlap.

## APPENDIX B: NaCl THIN FILM SAMPLES

The samples are NaCl microcrystalline thin films (Lebow) deposited on 30 nm thick Si$_3$N$_4$ membranes (Norcada NX5050X). NaCl thin films of 50 nm thickness were used in the studies reported here. A range of thin film thicknesses including 30 nm, 50 nm, and 70 nm, as well as a 120 nm film on a 50 nm substrate were initially tested to optimize experimental signal while avoiding propagation effects and signal reabsorption. The properties and purity of sub-micron thin films are difficult to measure – especially at sub-100 nm thicknesses. Lebow indicates the evaporant used during NaCl deposition is 99.9% pure and that such thin films are typically microcrystalline; therefore, it is very unlikely the films are single crystalline. A NaCl film thickness of 50 nm was found to provide a good compromise between signal strength and self-absorption. Dramatic spectral changes in films greater than 100 nm have been reported such as strong attenuation of absorption features [17,18]. The samples are mounted on a three-axis translation stage at the focal point of all three beams. The illuminated area was refreshed every minute using vertical and horizontal sample translation to avoid artifacts from sample heating and damage. The gas flow cell used for external calibrations and finding spatial and temporal overlap is dually mounted on the translation stage. Samples were received under N$_2$ purge and stored in a vacuum chamber to avoid sample contamination.